\documentclass{PoS}

\PoS{PoS(BDMH2004)100}
\usepackage{epsfig}

\def\simgt{\hbox{\rlap{\raise 0.425ex\hbox{$>$}}\lower 0.65ex\hbox{$\sim$}}}
\def\simlt{\hbox{\rlap{\raise 0.425ex\hbox{$<$}}\lower 0.65ex\hbox{$\sim$}}}

\def\simgt{\hbox{\rlap{\raise 0.425ex\hbox{$>$}}\lower 0.65ex\hbox{$\sim$}}}
\def\simlt{\hbox{\rlap{\raise 0.425ex\hbox{$<$}}\lower 0.65ex\hbox{$\sim$}}}

\def\grad{\phi_{\scriptscriptstyle\nabla}}
\def\radinx{\alpha}
\def\shear{\phi_\gamma}
\def\simgt{\hbox{\rlap{\raise 0.425ex\hbox{$>$}}\lower 0.65ex\hbox{$\sim$}}}
\def\simlt{\hbox{\rlap{\raise 0.425ex\hbox{$<$}}\lower 0.65ex\hbox{$\sim$}}}

\title{Strong Lensing Constraints on the Properties of Cluster Galaxies}

\ShortTitle{Substructure in Clusters}

\author{\speaker{Liliya L. R. Williams}\\
        Univ. of Minnesota, Minneapolis, MN, USA\\
        E-mail: \email{llrw@astro.umn.edu}}

\author{Prasenjit Saha\\
        Queen Mary \& Westfield, Univ. of London, UK\\
        E-mail: \email{p.saha@qmul.as.uk}}

\abstract{A recently discovered quadruply-imaged QSO, SDSS J1004+4112
(Inada et al. 2003; Oguri et al. 2004)
in the core of a $z=0.68$ galaxy cluster has an unprecedented image
separation of $\sim 13''$. This lens gives us a unique opportunity 
to study the detailed mass distribution in the central regions of this 
cluster. We present free-form 
reconstructions of the lens using recently developed methods. 
The projected mass within 100 kpc is well-constrained as 
$5\pm 1\, \times 10^{13} M_\odot$, consistent with previous simpler models. 
Unlike previous 
work, however, we are able to detect structures in the lens 
associated with cluster galaxies. We estimate the mass associated 
with these galaxies, and show that they contribute not more than 
about 10\% of the total cluster mass within 100 kpc. Typical galaxy 
masses, combined with typical luminosities yield a rough estimate 
of their mass-to-light ratio, which is $\simlt\, 10$, implying
that these galaxies consist mostly of stars, and possess little 
dark matter.}

\FullConference{Baryons in Dark Matter Halos\\
		 5-9 October 2004\\
		 Novigrad, Croatia}

\begin{document}

{\bf {Method.}}~ The lens is divided into 100-1000
independent mass pixels.
Image positions are taken as fixed model (primary) constraints.
Because these are greatly outnumbered
by the unknowns, secondary constraints are needed.
{\em PixeLens\/} generates a large number of individual mass maps; we
show the ensemble averages, or `best estimate' maps, from the Bayesian 
point of view.

Secondary Model Constraints, or Priors: 
(1) No pixel can have a negative mass: $\kappa\ge0$.
(2) The lens must be centrally concentrated.
{\em PixeLens\/} implements this by restricting
the direction of the mass density gradient, $\phi_\nabla$ at every 
pixel's location. The default value is $\phi_\nabla\le 45^\circ$,
meaning that the density gradient must point within $\pm 45^\circ$
of center. 
(3) Circularly averaged double logarithmic projected 
density slope in the image annulus, $\alpha$ 
(where $\rho_{2D}\propto r^{-\alpha}$) can be constrained.
(4) The influence of a nearby group or cluster 
can be incorporated as an external shear, whose approximate direction, 
$\phi_\gamma$, is a model input. A given input $\phi_\gamma$ allows any 
shear direction within $\pm 45^\circ$ of $\phi_\gamma$.

{\bf {Results.}}~ We show results using two sets of priors: 
{\it Prior A} has $\grad \leq 45^\circ$, while
{\it Prior B} has $\grad \leq 8^\circ$.
For both, $0.25\leq\radinx\leq3$, and $\shear=10^\circ\pm45^\circ.$ 
We experimented with several other types of priors. The basic results do not change:
(1) All recovered mass maps show the Northern and the South-Eastern galaxy
groupings;
(2) Fraction of cluster mass associated with individual galaxies $\simlt 10\%$;
(3) Typical $M/L$ per galaxy is $\simlt 15$.

\begin{figure}
\epsfig{file=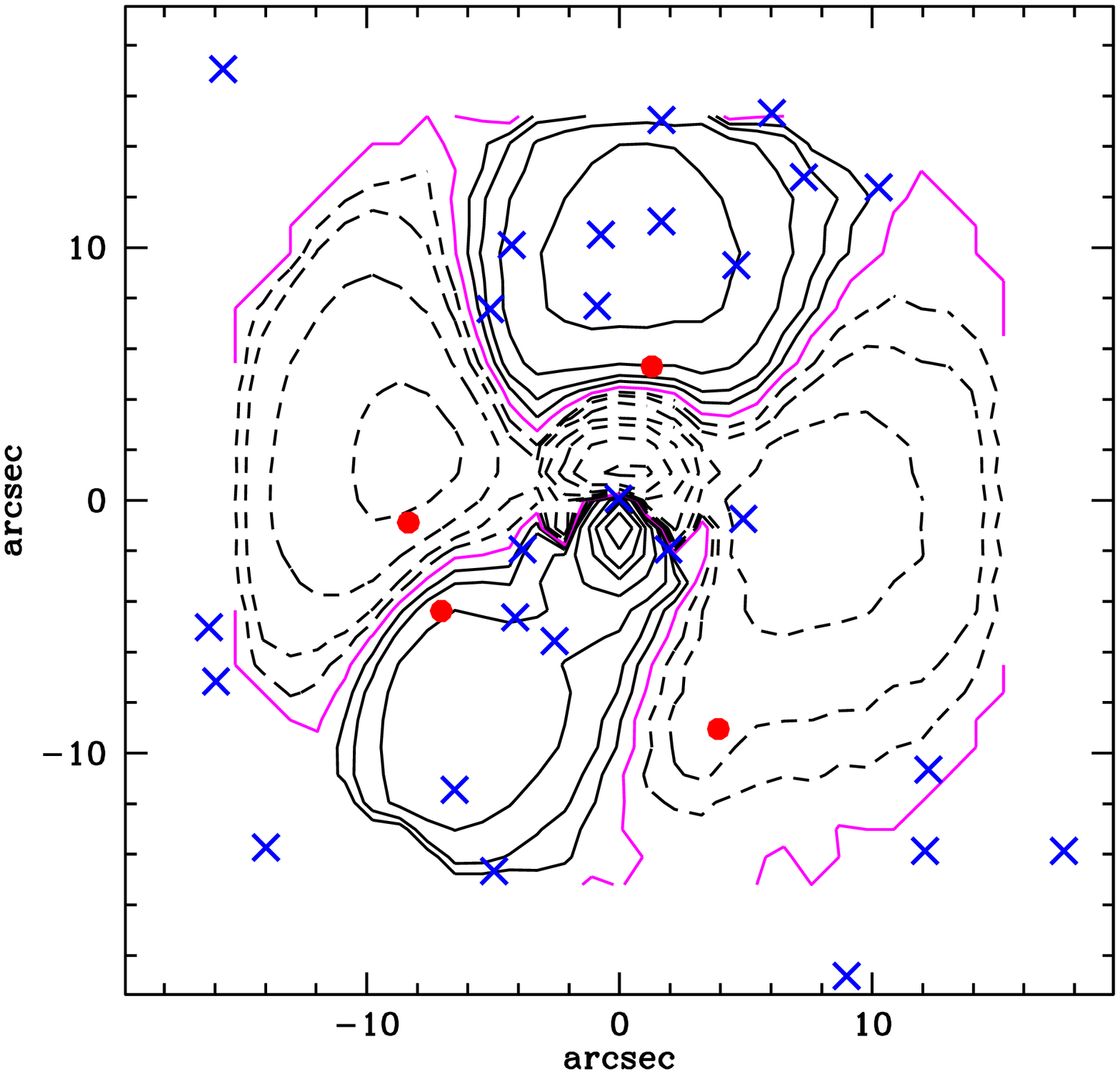, width=0.5\textwidth}
\epsfig{file=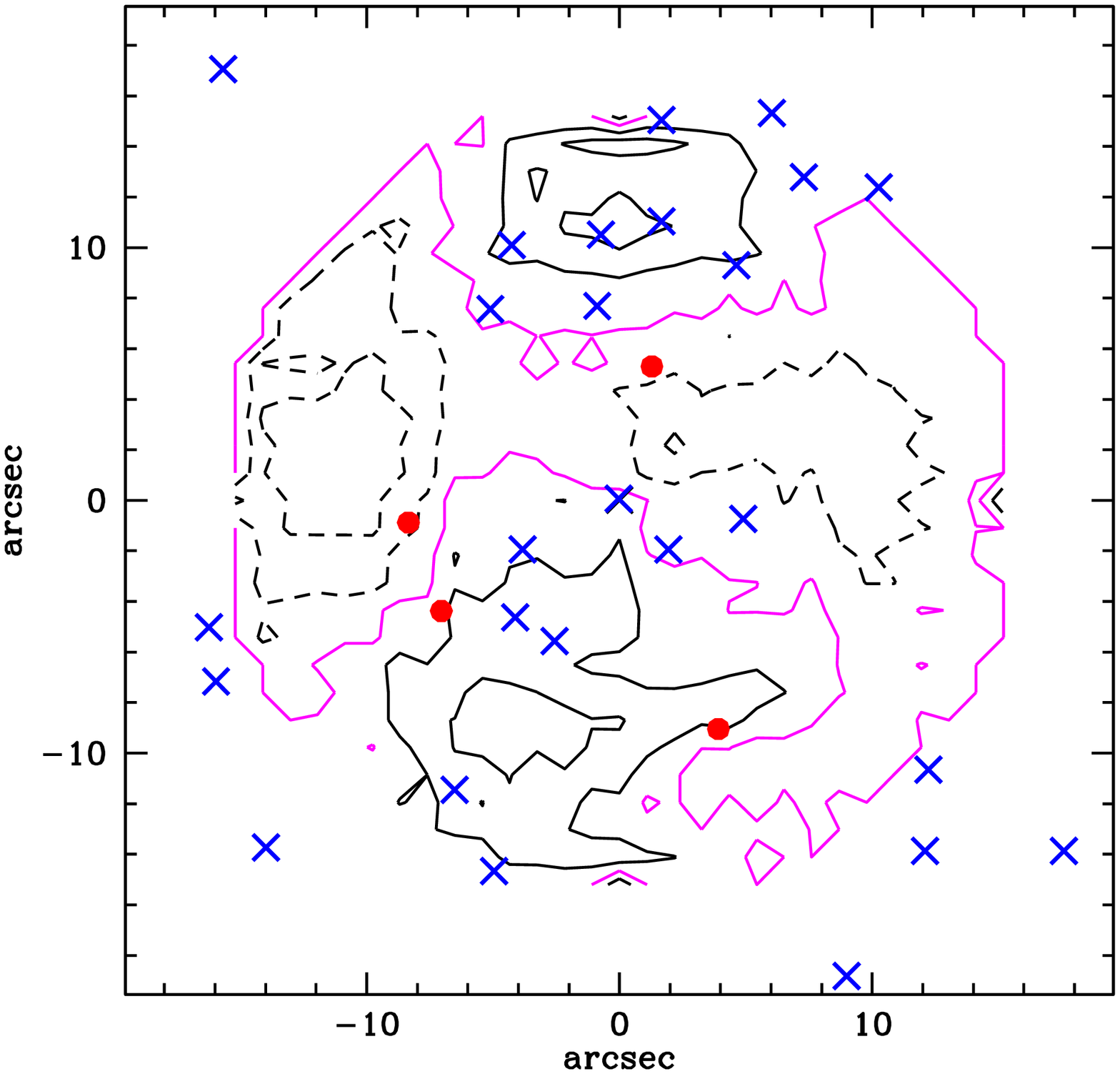, width=0.5\textwidth}
\caption{{\bf Residual mass maps} obtained by subtracting the circularly averaged mass
distribution from {\em PixeLens\/} ensemble average maps. The solid black contours 
indicate positive surface mass density residuals, and are drawn at
surface mass density, in terms of critical for lensing of $\kappa=0.025,0.05,0.1,\ldots$, 
or, equivalently,  $3.15, 6.3, 12.6,\ldots  M_\odot/\square''$. The dashed black
contours indicate negative surface mass density residuals. Purple contour indicates
zero density residual. The reconstruction
window has radius $15.2''$, the sky scale being $\simeq7.5\rm\,kpc/arcsec$;
the full horizontal scale is 300$h_{65}^{-1}$~kpc.
In each panel the mass maps shown are averages from ensembles of 500;
additional smoothing has been applied to the maps using $\sigma=0.67''$. 
Red dots are QSO images.
Blue crosses are galaxies with $i<24$, taken from Fig.~13a of Oguri et al. (2004). 
{\em Left:\/} Prior A. 
The derived typical galaxy M/L$\approx\!12$, while the fraction of mass 
associated with galaxies is $\sim\!9\%$.  
{\em Right:\/} Prior B. 
The derived typical galaxy M/L$\approx\!3$, and the fraction of mass 
associated with galaxies is $\sim\!2\%$ (see Williams \& Saha 2004 for details).}
\label{residual}
\end{figure}

\end{document}